\documentclass[twocolumn,showpacs,preprintnumbers,amsmath,amssymb,prl]{revtex4}

\usepackage{graphicx}
\usepackage{dcolumn}
\usepackage{bm}

\usepackage{mathptmx, courier, pifont}
\usepackage[scaled=0.92]{helvet}
\usepackage[T1]{fontenc}
\usepackage{textcomp}
\usepackage{color}
\usepackage{colordvi}

\newcommand{\quarterthin}{\kern 0.0417em}

\begin{document}


\title{Towards a unified realistic shell-model Hamiltonian with the
monopole-based universal force}

\author{K.~Kaneko$^{1}$, T.~Mizusaki$^{2}$,
Y.~Sun$^{3,4}$, S.~Tazaki$^{5}$ }

\affiliation{
$^{1}$Department of Physics, Kyushu Sangyo University, Fukuoka
813-8503, Japan \\
$^{2}$Institute of Natural Sciences, Senshu University, Tokyo
101-8425, Japan \\
$^{3}$Department of Physics and Astronomy, Shanghai Jiao Tong
University, Shanghai 200240, People's Republic of China \\
$^{4}$Institute of Modern Physics, Chinese Academy of Sciences,
Lanzhou 730000, People's Republic of China \\
$^{5}$Department of Applied Physics, Fukuoka University, Fukuoka
814-0180, Japan
}

\date{\today}

\begin{abstract}
We propose a unified realistic shell-model Hamiltonian employing the
pairing plus multipole Hamiltonian combined with the monopole
interaction constructed starting from the monopole-based universal
force by Otsuka {\it et al.} (Phys. Rev. Lett. 104, 012501 (2010)).
It is demonstrated that the proposed PMMU model can consistently
describe a large amount of spectroscopic data as well as binding
energies in the $pf$ and $pf_{5/2}g_{9/2}$ shell spaces, and could
serve as a practical shell model for even heavier mass regions.
\end{abstract}

\pacs{21.30.Fe, 21.60.Cs, 21.10.Dr, 27.50.+e}

\maketitle

The nuclear shell-model interaction can in principle be derived
microscopically from the free nucleon-nucleon force. In fact, such
attempts were made in the early years for the beginning of the shell
\cite{Kuo68,Jensen95}. However, it was soon after realized that such
an interaction fail to describe binding energies, excitation
spectra, and transitions if many valence nucleons were considered.
To reproduce experimental data, considerable efforts have been put
forward to construct the so-called effective interactions, such as
USD \cite{Brown88} and USDA/B \cite{Brown06} for the $sd$ shell,
KB3G \cite{Poves81} and GXPF1A \cite{Honma04} for the $pf$ shell,
and JUN45 \cite{Honma09} and jj4b \cite{jj4b} for the
$pf_{5/2}g_{9/2}$ model space. Each of these interactions is
applicable to a given model space while mutual relations among them
are obscure. On the other hand, it has been shown \cite{Dufour96}
that realistic effective interactions are dominated by the pairing
plus quadrupole-quadrupole ($P+QQ$) terms with the monopole
interaction. This finding not only makes the understanding of
effective interactions in nuclei intuitive, but may also be used to
unify effective interactions for different model spaces. In
particular, one may begin to talk about universality for shell
models.

Along the lines of this thought, we have carried out shell-model
calculations using the extended $P+QQ$ Hamiltonian combined with the
monopole terms ($EPQQM$), which are regarded as corrections for the
average monopole interaction. It has been demonstrated that despite
of its simplicity, this $EPQQM$ model works surprisingly well for
different mass regions, for example, the proton-rich $pf$ shell
\cite{Hasegawa01} and the $pf_{5/2}g_{9/2}$ shell \cite{Kaneko02},
the neutron-rich $fpg$ shell \cite{Kaneko08}, and the $sd$-$pf$
shell region \cite{Kaneko11}. It has also been successfully applied
to the neutron-rich nuclei around $^{132}$Sn \cite{Jin11,Wang13}.
However, these calculations rely heavily on phenomenological
adjustments on the monopole corrections. Consequently, the $EPQQM$
model cannot describe binding energies or unusual structures such as
the first excited $0^{+}$ state of Zn and Ge isotopes around $N=40$
\cite{Hasegawa07}, and it essentially provides no information about
the unconventional shell evolution in neutron-rich nuclei.

\begin{figure}[b]
\includegraphics[totalheight=8.0cm]{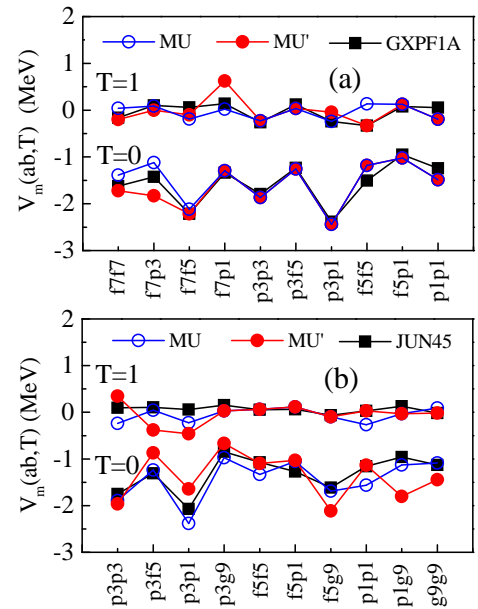}
\caption{(Color online) Monopole matrix elements for (a) $pf$ and
(b) $pf_{5/2}g_{9/2}$ model space. The original $V_{m}^{MU}$ and
modified $V_{m}^{MU'}$ are shown by open and full circles,
respectively. The monopole matrix elements of the GXPF1A interaction
\cite{Honma04} for $pf$ shell and JUN45 \cite{Honma09} for
$pf_{5/2}g_{9/2}$ shell are also shown for comparison. For orbital
labels, "f7p3", for example, stands for $a=f_{7/2}$ and
$b=p_{3/2}$.}
 \label{fig1}
\end{figure}

The monopole interaction is a crucial ingredient for successful
shell-model calculations. It is defined as \cite{PZ81}
\begin{eqnarray}
V_{m}(ab,T) & = & \frac{\sum_{J}(2J+1)V_{ab,ab}^{JT}}{\sum_{J}(2J+1)},
          \label{eq:1}
\end{eqnarray}
where $V_{ab,ab}^{JT}$ are the interaction matrix elements. The
connection between the monopole interaction and the tensor force
\cite{Otsuka01} was confirmed, which explains the shell evolution
\cite{Otsuka05}. It was shown \cite{Zuker03,Otsuka10a} that also
three-nucleon forces are important for the monopole interaction
between valence neutrons. Recently, novel general properties of the
monopole components in the effective interaction have been
demonstrated by Otsuka {\it et al.} by introducing the
monopole-based universal force $V_{MU}$ \cite{Otsuka10b}, which
consists of the Gaussian central force and the tensor force. The
proposed force has been successfully applied to light nuclei
\cite{Yuan12,Utsuno12}. As seen in Fig. \ref{fig1}, the monopole
matrix elements $V_{m}^{MU}$ obtained from this force are very
similar to those of the GXPF1A interaction \cite{Honma04} for the
$pf$ and JUN45 \cite{Honma09} for the $pf_{5/2}g_{9/2}$ shell space.
This let us to speculate that the $V_{MU}$ force proposed in Ref.
\cite{Otsuka10b} would be universal, maybe applicable to different
mass regions. Thus it is of great interest to investigate the
applicability of $V_{MU}$ by carrying out large-scale shell-model
calculations for nuclei in medium-mass regions. In this Rapid
Communication, we show that the monopole interaction derived from
the monopole-based unified force \cite{Otsuka10b}, with refitting
some monopole terms in accordance with each of the model spaces,
works well as an important part of a unified effective Hamiltonian
for the $pf$ and $pf_{5/2}g_{9/2}$ spaces. The conclusion is
supported by a systematical study of binding energies and by
performing detailed spectroscopic calculations for a wide range of
nuclei.

Our proposed Hamiltonian combines the pairing plus multipole terms
with the monopole interaction $V_{m}^{MU}$ (hereafter termed the
PMMU model)
\begin{eqnarray}\label{eq:2}
 H & = & H_{0} + H_{PM}  + H_{m}^{MU},  \\ \nonumber \\
 H_{0} & = & \sum_{\alpha} \varepsilon_a c_\alpha^\dag c_\alpha, \nonumber \\
 H_{PM}  & = &
 -  \sum_{J=0,2} \frac{1}{2} g_J \sum_{M\kappa} P^\dag_{JM1\kappa} P_{JM1\kappa} \nonumber \\
   && -  \frac{1}{2} \chi_2 \sum_M :Q^\dag_{2M} Q_{2M}:
    - \frac{1}{2} \chi_3 \sum_M :O^\dag_{3M} O_{3M}:  \nonumber \\
 H_{m}^{MU}  & = &  \sum_{a \leq b, T} V_{m}^{MU}(ab,T)
 \sum_{JMK}A^\dagger_{JMTK}(ab) A_{JMTK}(ab). \nonumber
\end{eqnarray}
For the pairing plus multipole part, we take the $J=0$, 2 terms in
the particle-particle channel and the quadrupole-quadrupole ($QQ$)
and octupole-octupole ($OO$) terms in the particle-hole channel
\cite{Hasegawa01,Kaneko02}. (Higher order pairing and multipole
terms can be added if necessary.) We adopt the monopole matrix
elements constructed from the monopole-based universal force
\cite{Otsuka10b}, in which the Gaussian parameter is fixed as
$\mu=1.0$ fm, and the strength parameters used here are the same as
those in Ref. \cite{Otsuka10b}. The harmonic oscillator basis with
$\hbar \omega = 41 A^{-1/3}$ is used in the calculation of the
matrix elements. Note that in the monopole Hamiltonian given above,
$V_{m}^{MU}(ab,T)$ is the modified one labeled as $MU'$ in Fig.
\ref{fig1}. It should be emphasized that the monopole interaction
$V_{m}^{MU}$ is a part of the effective interaction in the present
model, not just a correction as treated in our previous papers (see,
for example, Ref. \cite{Hasegawa07}). In this way, effective
single-particle energies are crucially improved to reflect the shell
evolution.

\begin{figure*}[t]
\begin{center}
\includegraphics[totalheight=6.0cm]{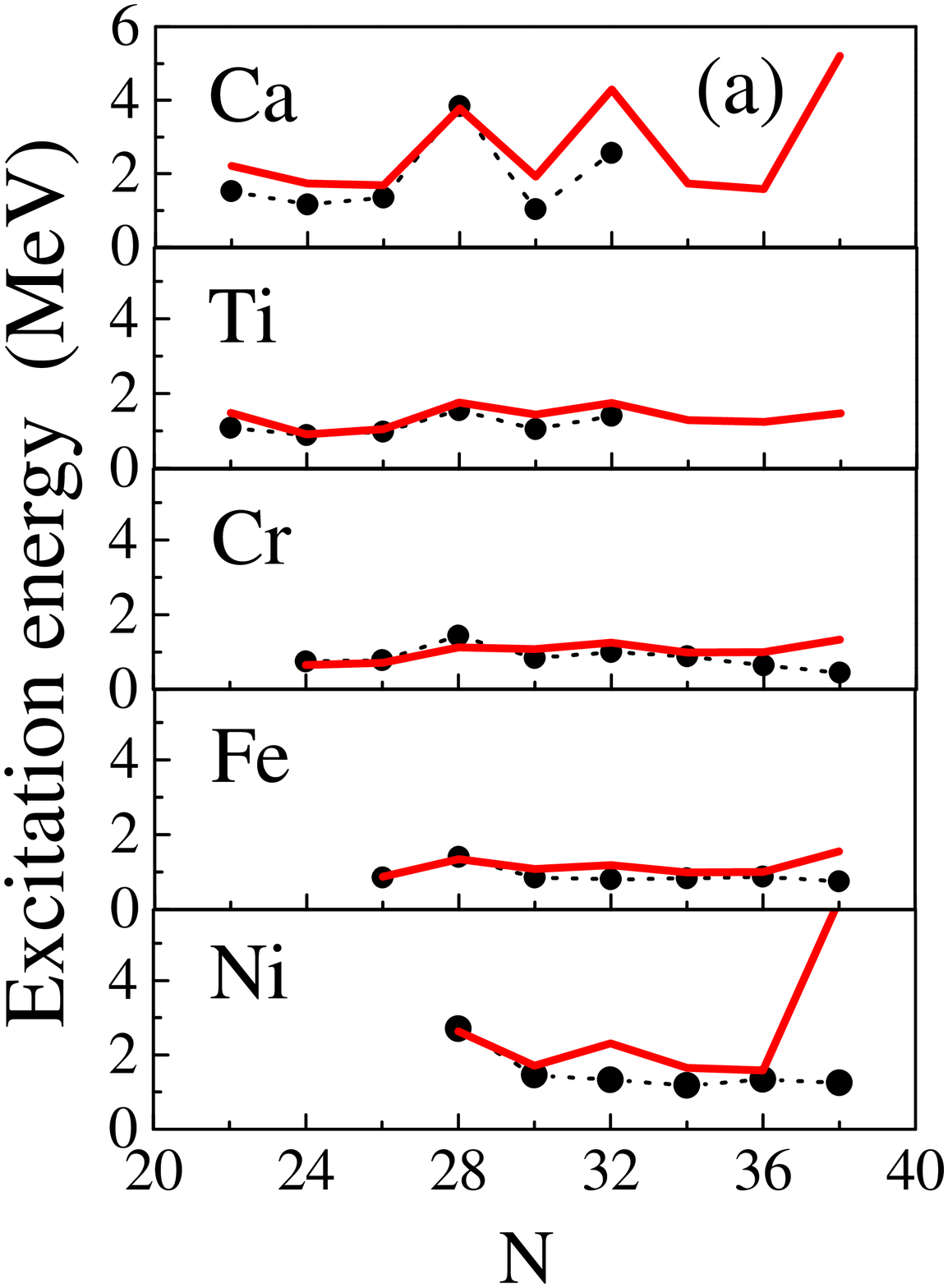}
\hspace{0.5cm}
\includegraphics[totalheight=6.0cm]{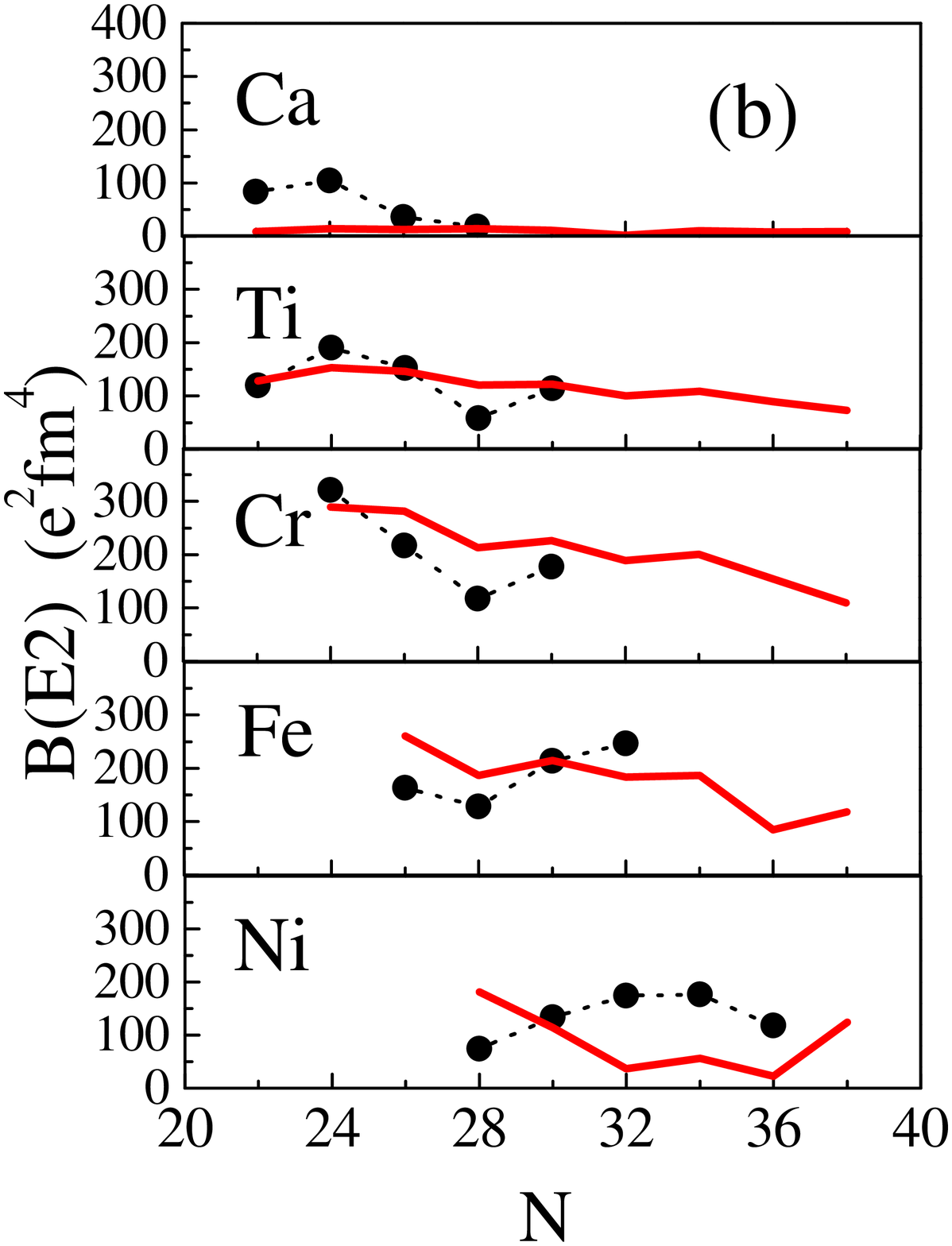}
\hspace{0.5cm}
\includegraphics[totalheight=6.0cm]{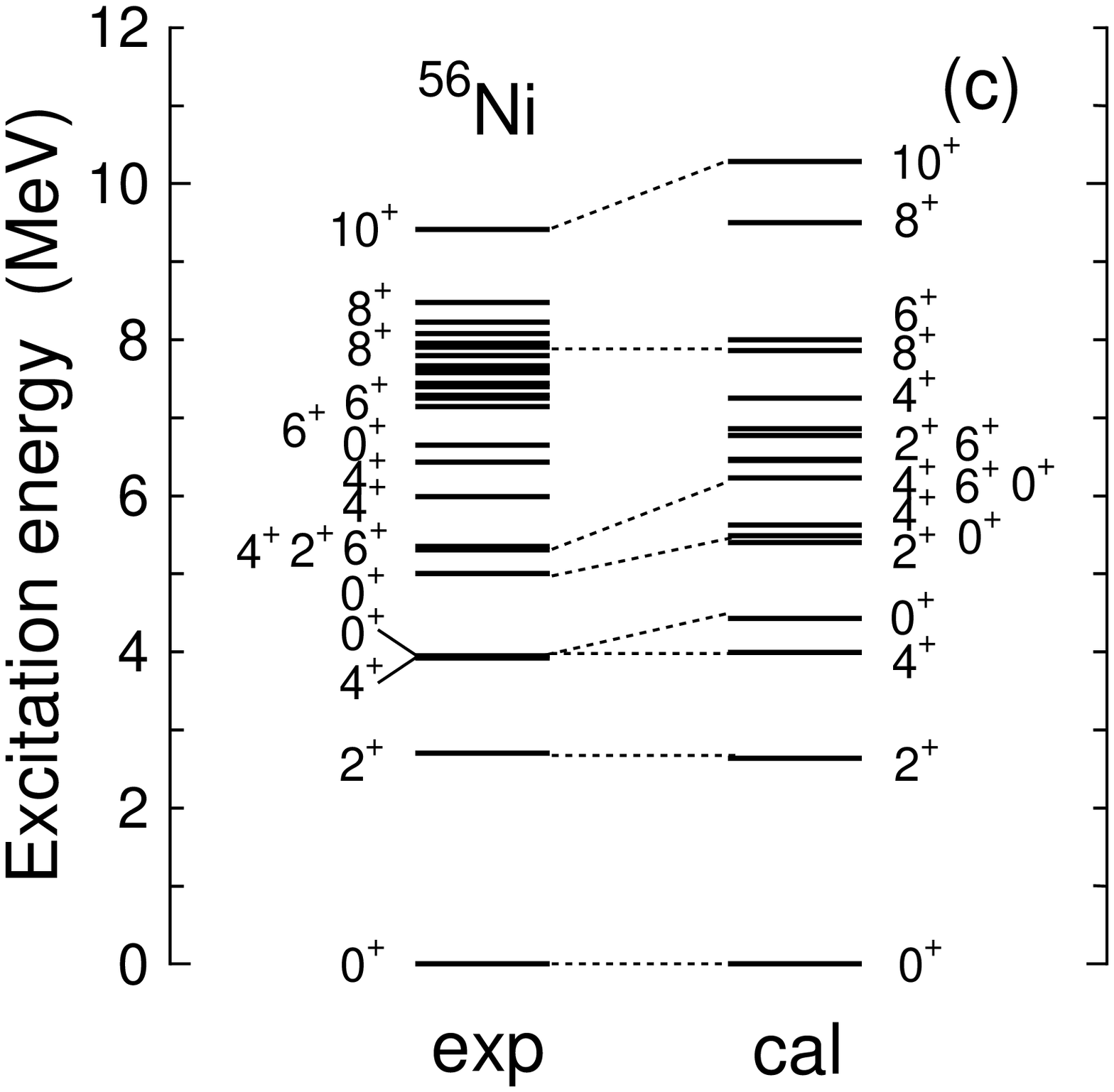}
\caption{(Color online) Systematics for $pf$-shell nuclei. (Left
panel) $2_{1}^{+}$ energy, (middle panel) $B(E2;2_{1}^{+}\rightarrow
0_{1}^{+})$ value, and (right panel) energy spectrum for $^{56}$Ni.
Filled circles and solid lines in the left and middle panels
indicate experimental data and the calculated results, respectively.
The experimental data are taken from \cite{ENSDF}.}
  \label{fig2}
 \end{center}
\end{figure*}
Shell-model calculations are performed with the code MSHELL64
\cite{Mizusaki}. We first apply the PMMU model in the $pf$ model
space.  The calculated total binding energy is obtained by adding an
appropriate Coulomb energy to the shell-model ground-state energy.
In the $pf$-shell calculation, Coulomb energies are evaluated as in
Ref. \cite{Langanke95} by using an empirical formula
$E_{C}=V_{pp}p(p-1)+V_{pn}pn+e_{p}p$, where $p$ ($n$) denotes the
number of valence protons (neutrons) and the adopted parameters are
$V_{pp}=0.10$, $V_{pn}=-0.203$, and $e_{p}=6.90$ (all in MeV). Data
from 95 nuclei \cite{Audi95}, $^{42-49}$Ca, $^{43-54}$Sc,
$^{48-55}$Ti, $^{50-60}$V, $^{51-62}$Cr, $^{52-63}$Mn, $^{53-64}$Fe,
$^{54-64}$Co, and $^{56-64}$Ni, are taken for fitting. All the
results for Ca, Sc, Ti, and Cr isotopes are obtained without any
truncation. For the other isotopes, the maximal number of allowed
particles  excited from the $f_{7/2}$ orbital is limited. The rms
deviation for binding energies is 707 keV. As a result, the
interaction strengths in (\ref{eq:2}) are determined to be
$g_{0}=20.2/A$, $g_{2}=228.2/A^{5/3}$, $\chi_{2}=228.2/A^{5/3}$, and
$\chi_{3}=0.0$, and the single-particle energies to be
$\varepsilon_{f7/2} =-7.78$, $\varepsilon_{p3/2} =-5.68$,
$\varepsilon_{f5/2} =-1.28$, and $\varepsilon_{p1/2} =-3.88$ (all in
MeV). For the $pf$ shell, 10 monopole terms are also modified for
fitting. The results are displayed as $V_{m}^{MU'}$ in Fig.
\ref{fig1}(a). All the monopole matrix elements are scaled with a
factor $(42/A)^{0.3}$ in the calculation.

Our strength parameters $g_{0}$ and $\chi_{2}$ agree, respectively,
with $g_{0}=20/A$ and $\chi_{2}=240/A^{5/3}$ of Bes and Sorensen
\cite{Bes69}. We may also compare our strengths with those derived
from the GXPF1A interaction using the prescription of Dufour and
Zuker \cite{Dufour96}. They can be estimated from the following
equations; $g_{0} = |E^{01}|\frac{\hbar \omega}{\hbar
\omega_{0}}\Omega_{r}^{-1}$ and $\chi_{2} = 2|e^{20}|\frac{\hbar
\omega}{\hbar \omega_{0}}N_{r}^{-2}$, where $E^{JT}$ and $e^{JT}$
are, respectively, $E$- and $e$-eigenvalues, and $\Omega_{r}=0.655
A^{2/3}$, $N_{r}^{2}=0.085 A^{4/3}$, $\hbar \omega=40A^{-1/3}$, and
$\hbar \omega_{0}=9.0$ (see Ref. \cite{Dufour96}). The
$E$-eigenvalue of the interaction matrix for $J=0,T=1$ is
$E^{01}=-4.18$, and the $e$-eigenvalue for $J=2,T=0$ particle-hole
channel is $e^{20}=-2.92$ \cite{Honma04}. The obtained strength
parameters using the above values are $g_{0}=28.4/A$ and
$\chi_{2}=305.4/A^{5/3}$. A comparison of these values with the
strength parameters determined from our calculation reveals that
both have the same mass dependence while ours are about 0.73 times
smaller than those of GXPF1A. This result is interesting because it
indicates that the GXPF1A model and our PMMU model for the $pf$
shell are different, but may be closely related.

In Fig. \ref{fig2}, the results from calculations with the PMMU
Hamiltonian are compared with available experimental data for the
$pf$ shell nuclei. The left panel of Fig. \ref{fig2} shows
excitation energies of the $2_{1}^{+}$ state for Ca, Ti, Cr, Fe, and
Ni isotopes. The systematic behavior of the $2_{1}^{+}$ energy
levels is reasonably described for these isotopes although the
calculation produces a small jump corresponding to the shell closure
at $N=32$ for the Ca and Ni isotopes. At $N=40$, deviations from the
experimental levels are seen. Especially, the calculated $2_{1}^{+}$
for $^{58}$Ca and $^{66}$Ni are quite large, which may indicate that
the $g_{9/2}$ orbital could be important for these levels. In the
middle panel, calculated $B(E2;2_{1}^{+}\rightarrow 0_{1}^{+})$
values are shown, for which the effective charges are taken as
$e_{p}=1.5e$ and $e_{n}=0.5e$. The calculation basically accounts
for the variation trend, however with deviations from data for the
beginning and end of the shell. For Ca nuclei with $N \le 24$ and Ni
nuclei with $N \ge 32$, the calculated $B(E2)$ values are smaller
than experiment. The reason for the deviations is understandable.
Those in Ca may suggest that the contribution of core excitations
from the $sd$ shell is large and those in Ni may indicate that
inclusion of the $g_{9/2}$ and $d_{5/2}$ orbitals in the model space
is important. As an example of detailed spectroscopic calculations,
the obtained energy levels for $^{56}$Ni are compared with data in
the right panel. It can be seen that the calculation reproduces the
data very well. In particular, the agreement with the experimental
levels below 5 MeV is excellent. Experimental information for
electromagnetic transitions is rather limited for this nucleus. The
calculated values $B(E2;2_{1}^{+}\rightarrow 0_{1}^{+})=181.2$
$e^{2}$fm$^{2}$ and $B(E2;4_{1}^{+}\rightarrow 2_{1}^{+})=121.0$
$e^{2}$fm$^{2}$ are compared with experimental ones \cite{Kraus94}
120$\pm$24 $e^{2}$fm$^{2}$ and $<305$ $e^{2}$fm$^{2}$, respectively.

\begin{figure*}[t]
 \begin{center}
\includegraphics[totalheight=6.0cm]{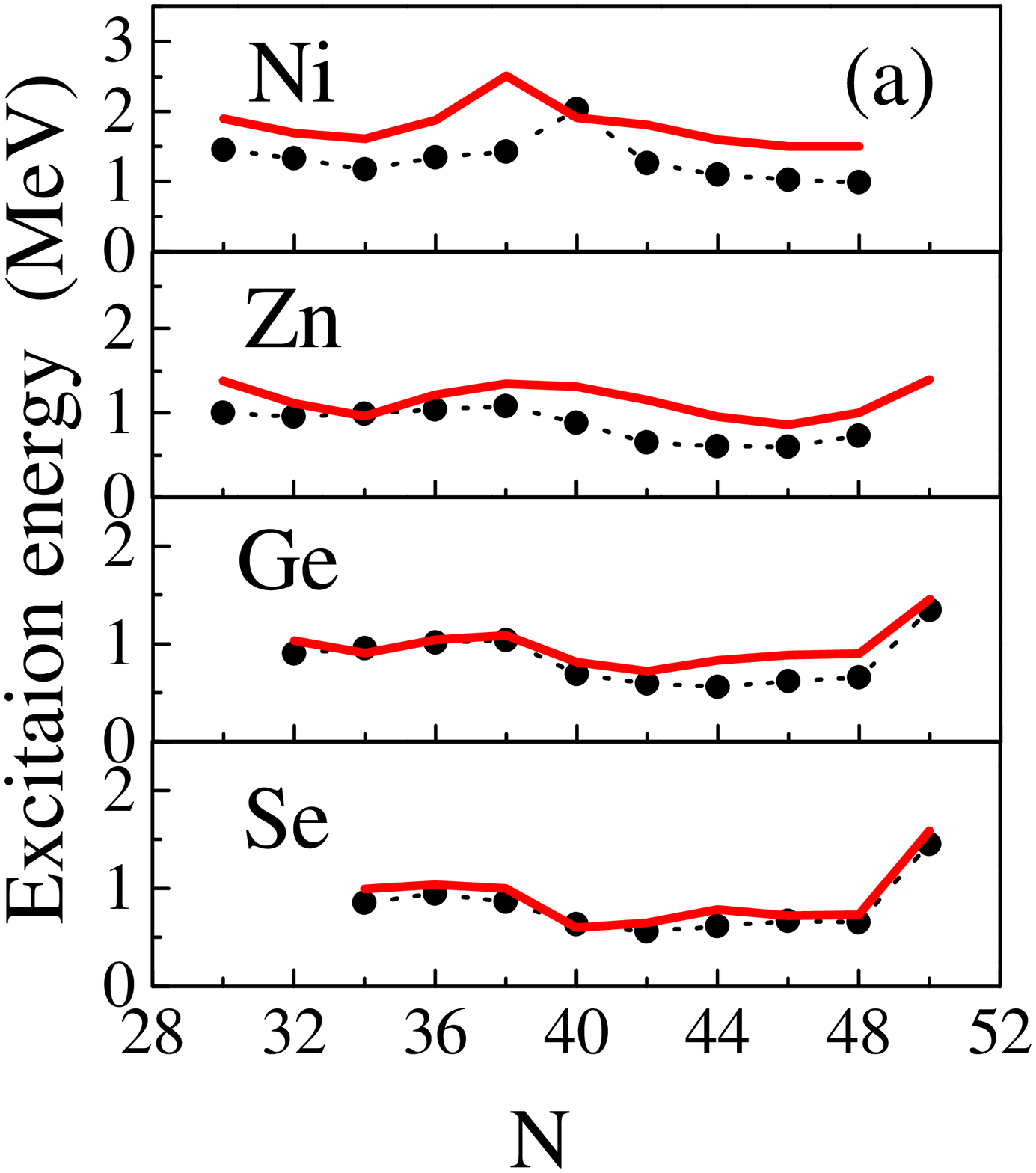}
\hspace{0.2cm}
\includegraphics[totalheight=6.0cm]{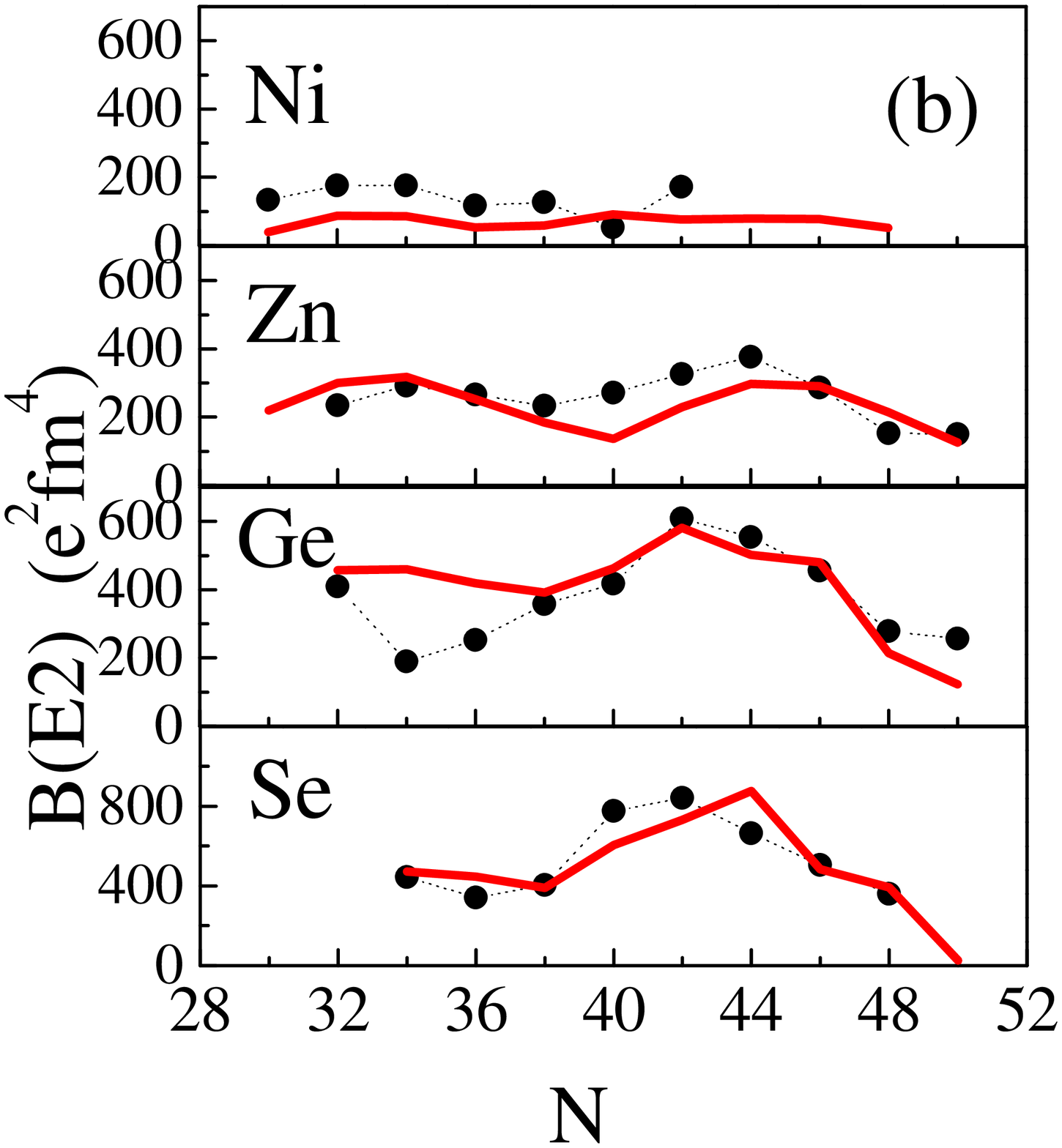}
\hspace{0.2cm}
\includegraphics[totalheight=6.0cm]{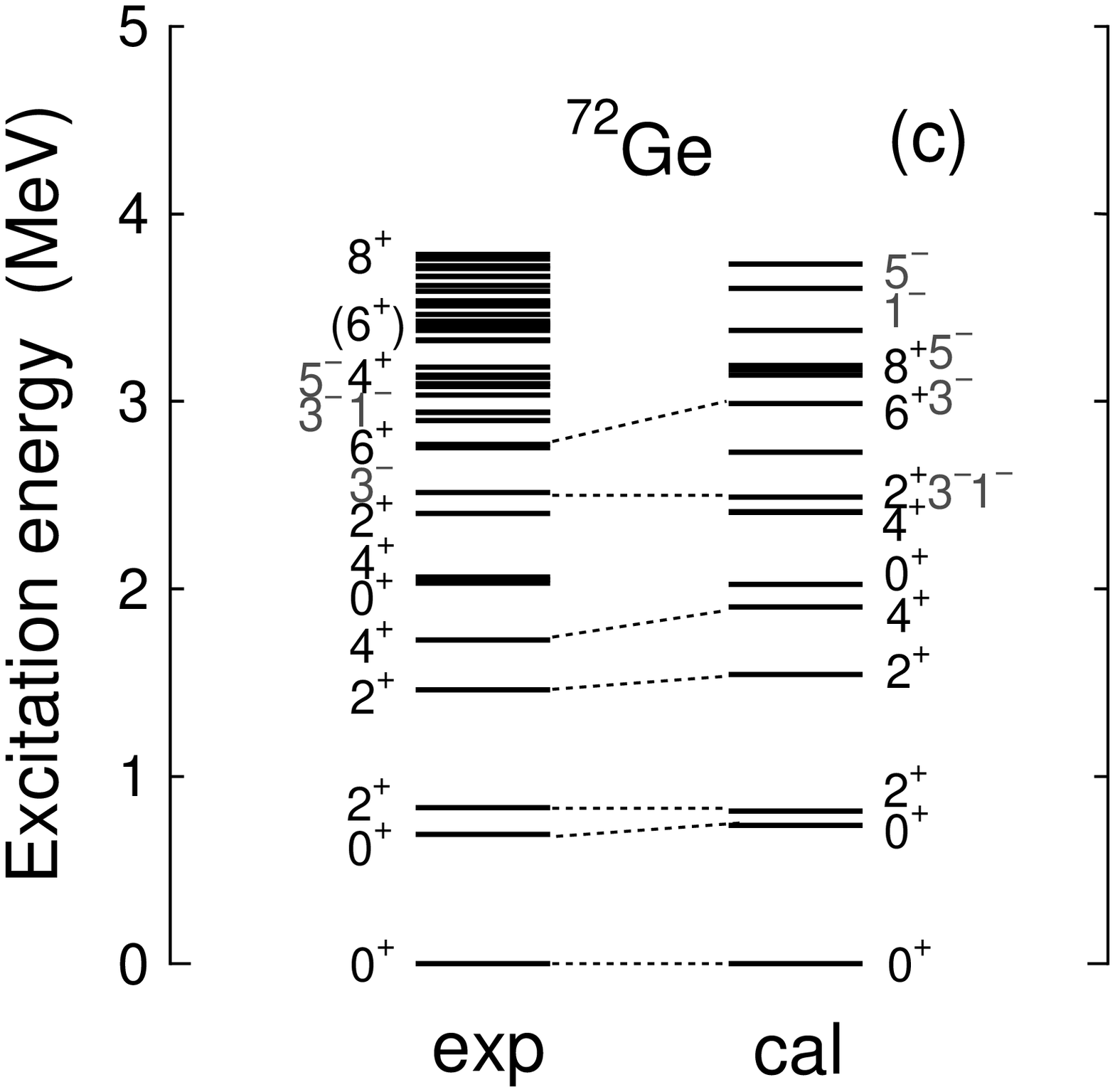}
\caption{(Color online) Systematics for $pf_{5/2}g_{9/2}$-shell
nuclei. (Left panel) $2_{1}^{+}$ energy, (middle panel)
$B(E2;2_{1}^{+}\rightarrow 0_{1}^{+})$ value, and (right panel)
energy spectrum for $^{72}$Ge. Filled circles and solid lines in the
left and middle panels indicate experimental data and the calculated
results, respectively. The experimental data are taken from
\cite{ENSDF}.}
  \label{fig3}
 \end{center}
\end{figure*}
\begin{figure*}[t]
\includegraphics[totalheight=6.0cm]{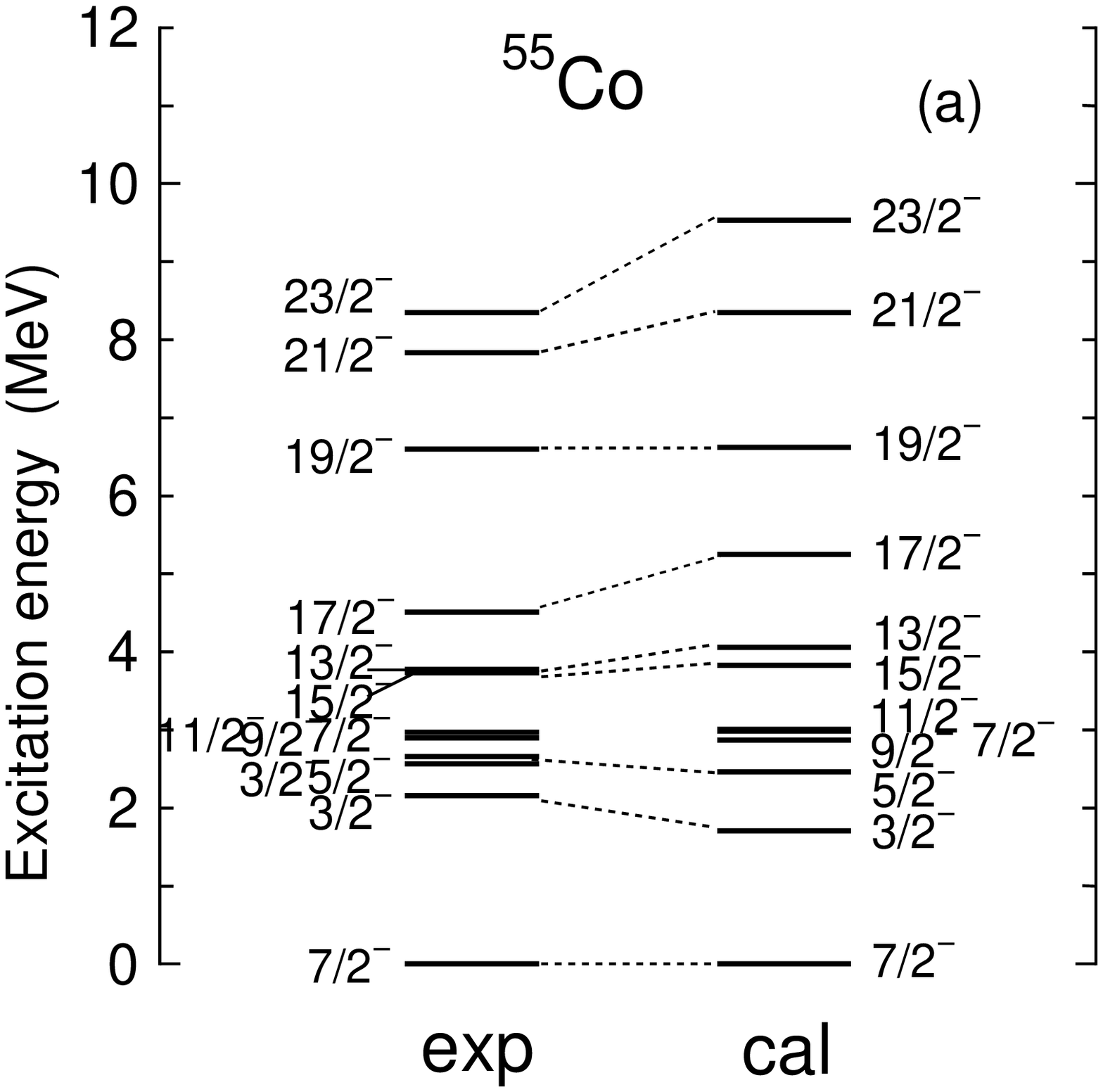}
\hspace{2cm}
\includegraphics[totalheight=6.0cm]{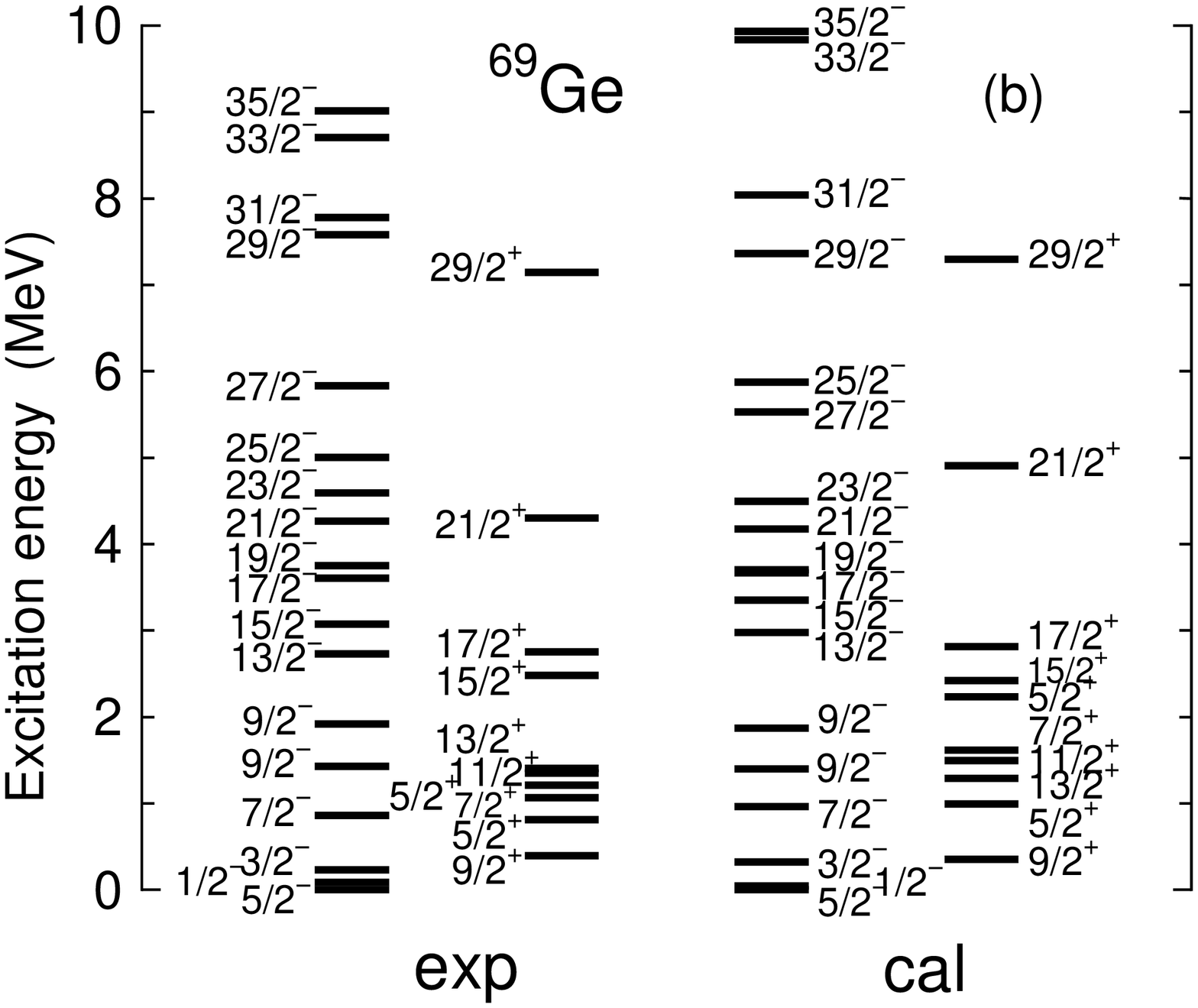}
\caption{Calculated energy levels for $^{55}$Co and $^{69}$Ge are
compared with the corresponding experimental data taken from
\cite{ENSDF}.}
  \label{fig4}
\end{figure*}

Next we apply the PMMU model for the $pf_{5/2}g_{9/2}$-space.
Binding energies are evaluated by using the empirical formula for
the Coulomb energy with the parameter set 2 given in Table I of Ref.
\cite{Cole99}. The data from 91 nuclei, $^{64-76}$Ni, $^{64-78}$Cu,
$^{65-80}$Zn, $^{66-80}$Ga, $^{69-80}$Ge, $^{67-78}$As, and
$^{73-80}$Se \cite{Audi95}, are taken for the fitting calculation.
For Ni, Cu, Zn, Ga, and Ge isotopes, the calculation is performed
without any truncation, while for As and Se isotopes, some
truncations are introduced. The rms deviation for binding energies
is 691 keV. The resulting interaction strengths in Eq. (\ref{eq:2})
are $g_{0}=18.0/A$, $g_{2}=0.0$, $\chi_{2}=334.0/A^{5/3}$, and
$\chi_{3}=259.2/A^{2}$, and the single-particle energies are
$\varepsilon_{p3/2} =-9.40$, $\varepsilon_{f5/2} =-8.29$,
$\varepsilon_{p1/2} =-7.49$, and $\varepsilon_{g9/2} =-5.70$ (all in
MeV). For this shell space, 14 monopole terms are modified for
fitting, with the results displayed as $V_{m}^{MU'}$ in Fig.
\ref{fig1}(b). All the monopole matrix elements are scaled with a
factor $(58/A)^{0.3}$ in the calculation.

In Fig. \ref{fig3}, we show the results for Ni, Zn, Ge, and Se
isotopes.  $E_x(2_{1}^{+})$ and $B(E2;2_{1}^{+}\rightarrow
0_{1}^{+})$ values are compared with experimental data in the left
and middle panel, respectively. The effective charges are taken as
$e_{p}=1.5e$ and $e_{n}=1.1e$. For Zn isotopes, both
$E_x(2_{1}^{+})$ and $B(E2)$ are described reasonably well. Near
$N=40$, the calculated $B(E2)$ values nevertheless underestimate the
data. The calculation also reproduces well the experimental data for
Ge and Se isotopes. The feature that $B(E2)$ values for both
isotopic chains increase around $N=44$ is correctly described. For
Ge isotopes, the overall agreement in $E_x(2_{1}^{+})$ is obtained.
The calculated $B(E2)$ values are however found too large for
$^{66,68}$Ge, which was also seen in the JUN45 calculation
\cite{Honma09}. It was pointed out \cite{Luttke12} that these
experimental $B(E2)$ values can be reproduced if a smaller neutron
effective charge $e_{n}=0.5e$ is used. Thus this problem currently
remains as an open question. Finally for Ni isotopes, the observed
$2_{1}^{+}$ energy shows the largest value at $N=40$ while the
calculation suggests the largest at $N=38$. For all the Ni isotopes
the calculated $E_x(2_{1}^{+})$ are somewhat higher than experiment
and the $B(E2)$ values are smaller. The problem with the present
calculation may be the restriction in the $pf_{5/2}g_{9/2}$-shell
space without inclusion of the $f_{7/2}$ and $d_{5/2}$ orbitals.

$^{72}$Ge is an interesting nucleus because its first excited
$0^{+}$ state is found to be the lowest in all the Ge isotopes,
which lies below $2_{1}^{+}$. As one can see from the right panel of
Fig. \ref{fig3}, this feature is correctly reproduced by the present
calculation. It is remarkable that the calculation shows an
one-to-one correspondence with experimental levels for
positive-party states up to high excitations. The calculated
$B(E2;2_{1}^{+}\rightarrow 0_{1}^{+})=462.0$ $e^{2}$fm$^{2}$ and
$B(E2;4_{1}^{+}\rightarrow 2_{1}^{+})=712.1$ $e^{2}$fm$^{2}$ are
compared well with the experimental data 418.1$\pm$72
$e^{2}$fm$^{2}$ and 658.3$\pm$88 $e^{2}$fm$^{2}$ \cite{ENSDF},
respectively.

Description of odd-mass nuclei is generally a challenging task for
shell-model calculations. To further test the PMMU model, we take
$^{55}$Co and $^{69}$Ge as examples for the $pf$ and
$pf_{5/2}g_{9/2}$ shell spaces, respectively. As one can see in Fig.
\ref{fig4}, the present model works well also for odd-mass nuclei.
If one assumes an inert $^{56}$Ni core, the low-lying states of
$^{55}$Co may be naively understood so as to be built by the simple
$f_{7/2}$ proton-hole configuration, having $7/2^{-}$ as the ground
state. However, our calculation indicates that for the yrast states
up to $17/2^{-}$, the $^{56}$Ni core is strongly broken and
excitations from the $f_{7/2}$ orbital to the upper $pf$-shell are
large. The higher spin $13/2^{-}$, $15/2^{-}$, and $17/2^{-}$ states
are described mainly by neutron excitations. For $^{69}$Ge,
agreement of the calculation with data is excellent for both
negative- and positive-parity states. In particular, the calculation
reproduces well the low-lying states including the ground state
$5/2^{-}$. Discrepancies in the $^{69}$Ge calculation are found for
the $25/2^{-}$, $33/2^{-}$, and $35/2^{-}$ states, for which the
theoretical levels lie by about 1 MeV higher than the experimental
data.

Although the present PMMU is constructed very differently from the
GXPF1A or JUN45 shell model, the striking similarities in their
monopole matrix elements shown in Fig. \ref{fig1} ensure that they
may describe shell structures equally well. We have found that the
monopole matrix elements in USD and USDA/B are also quite similar to
those in PMMU. However, the merit of PMMU is that its interaction
takes a very simple form with a common mass dependence for different
mass regions and has a much smaller number of parameters in the
Hamiltonian. One may expect that the model works even better for
heavier nuclei because the $P+QQ$ type of forces is well justified
there. This may open a promising path for shell models to be
consistently applied to heavy systems, for example, the $A\sim 100$
mass region.

In conclusion, we have made a step towards a unification of
effective shell-model Hamiltonian applicable to different mass
regions. The Hamiltonian adopts the well-established
pairing-plus-multipole force and is crucially combined with the
monopole interaction derived from the recently-proposed
monopole-based universal force \cite{Otsuka10b}. The constructed
PMMU model has been applied to a large number of nuclei in the $pf$
and $pf_{5/2}g_{9/2}$ shell regions. For both regions, the PMMU
model, with the resulting force strengths and the scaling parameters
with a common mass dependence adjusted to systematical reproduction
of binding energies, can well describe the experimental spectra and
transitions, for both the near-ground-state region and high-spin
excitations. It has been suggested that the PMMU model could be a
feasible method to unify different shell models so that one may
extend shell-model calculations to heavier systems.

Research at SJTU was supported by the National Natural Science
Foundation of China (Nos. 11135005 and 11075103) and by the 973
Program of China (No. 2013CB834401).



\end{document}